\documentclass[aps,prl,showpacs,twocolumn,superscriptaddress]{revtex4}

\usepackage{graphicx}
\usepackage{longtable}
\usepackage{amssymb}
\usepackage[ansinew]{inputenc}

\begin{document}

\bibliographystyle{prsty}

\title{Dynamic Jahn-Teller effect in electron transport through single C$_{60}$ molecules}
\author{T.~Frederiksen}
\email{thomas_frederiksen@ehu.es}
\affiliation{\mbox{Donostia International Physics Center (DIPC), Manuel de Landizabal Pasealekua 4, 
E-20018 Donostia-San Sebasti\'an, Spain}}
\affiliation{CIC nanoGUNE Consolider, Mikeletegi Pasealekua 56, E-20009 Donostia-San Sebasti\'an, Spain}
\author{K.~J.~Franke}
\affiliation{\mbox{Institut f\"ur Experimentalphysik, Freie Universit\"at Berlin, Arnimallee 14, D-14195 Berlin, Germany}}
\author{A.~Arnau}
\affiliation{\mbox{Donostia International Physics Center (DIPC), Manuel de Landizabal Pasealekua 4, 
E-20018 Donostia-San Sebasti\'an, Spain}}
\affiliation{\mbox{Depto.~de F\'isica de Materiales UPV/EHU, Facultad de Qu\'imica, Apartado 1072, 
E-20080 Donostia-San Sebasti\'an, Spain}}
\affiliation{\mbox{Centro de F\'isica de Materiales, Centro Mixto CSIC-UPV/EHU, 
E-20080 Donostia-San Sebasti\'an, Spain}}
\author{G.~Schulze}
\affiliation{\mbox{Institut f\"ur Experimentalphysik, Freie Universit\"at Berlin, Arnimallee 14, D-14195 Berlin, Germany}}
\author{J.~I.~Pascual}
\affiliation{\mbox{Institut f\"ur Experimentalphysik, Freie Universit\"at Berlin, Arnimallee 14, D-14195 Berlin, Germany}}
\author{N.~Lorente}
\affiliation{Centre d'Investigaci\'o en Nanoci\`encia i Nanotecnologia (CSIC-ICN),
Campus de la Universitat Aut\`onoma de Barcelona, E-08193 Bellaterra, Spain}

\date{April 21, 2008}


\begin{abstract}
Scanning tunneling spectra on single C$_{60}$ molecules that are sufficiently decoupled from the substrate exhibit a characteristic fine
structure, which is explained as due to the dynamic Jahn-Teller effect. Using electron-phonon couplings extracted from density
functional theory we calculate the tunneling spectrum through the C$_{60}^-$ anionic state and find excellent
agreement with measured data.
\end{abstract}

\maketitle

Orbital electronic degeneracy and stability of the nuclear configuration are incompatible unless the atoms of a 
molecule lie on a straight line. This statement is due to Jahn and Teller \cite{JT.1937}, who proved its general validity
in 1937 using group theory \cite{noteA}.
The theorem has important implications for non-linear molecules
with a degenerate electronic ground state, particularly when the degenerate electrons participate in the
binding of the molecule.
In such cases the molecule undergoes a structural distortion which lifts the electronic 
degeneracy by reducing the symmetry of the nuclear configuration, known as the static Jahn-Teller (JT) effect.
In certain situations \emph{several} distortions can lower the symmetry of the molecule, and hence lift the
electronic degeneracy. When such distorted states are also degenerate, the system will fluctuate between these 
equivalent configurations by quantum tunneling, resulting in pseudorotations \cite{PRB2006.Hands} and 
restoration of the parent symmetry, denoted the dynamic JT effect \cite{dresselhaus.95,PRB1994.Wang}.

The buckminsterfullerene C$_{60}$ is an exceptionally symmetric molecule, and its symmetry governs many of its
physical and chemical properties. In group theory it is classified by the icosahedral point group $I_h$, the group 
with the highest number of symmetry operations in three dimensions \cite{dresselhaus.95}. The geometrical symmetry
is also reflected in the electronic and vibrational properties, which are both highly degenerate. 
For instance, the highest occupied molecular orbital (HOMO) and the lowest unoccupied molecular 
orbital (LUMO) are orbitally fivefold and threefold degenerate, respectively. 
By adding electrons or holes to C$_{60}$ these degeneracies are explored and, hence, JT physics 
is expected to play an important r\^ole \cite{Manini-Dunn}.

Indeed, the JT effect has been reported in photoemission (PE) spectra of C$_{60}$ in the gas phase for 
different charge states of the molecule 
\cite{PRL1995.Gunnarsson,CPL1997.Bruhwiler,PRL2002.Canton,PRL2003.Manini,PRL2005.Tomita}.
Here, a fine structure in the intensity spectrum has been explained as originating in the underlying 
electron-phonon (e-ph) coupled problem. Signatures of JT have also been found in luminescence 
spectra of C$_{60}$ nanocrystals \cite{PRL2005.Cavar}. A more direct consequence of the JT effect has
been observed in low-temperature scanning tunneling microscopy (STM) images of potassium-doped 
fullerenes (K$_{3}$C$_{60}$) on Au(111), revealing a significant static deformation of the molecular
structure \cite{SCI2005.Wachowiak}. While vibronic effects have also been studied in the transport 
properties of C$_{60}$ molecules \cite{Nat2000.Park,JCP2002.Pascual,JPCB2005.Pradhan}, the dynamic
JT effect in single-molecule conductance has not been reported so far, presumably due to the 
difficulty of decoupling a molecule sufficiently from its environment. 

From the theoretical side there has been a substantial interest in describing the 
e-ph coupled problem in C$_{60}$, primarily motivated by the discovery of the superconducting 
phase of alkali doped fullerides A$_3$C$_{60}$ and its relation to the ``conventional''
e-ph mechanism \cite{Sci1991.Varma,PRB1993.Antropov,RMP1997.Gunnarsson}.
An accurate calculation of the e-ph coupling constants is fundamental to such descriptions.
In comparison with couplings fitted to photoemission data on C$_{60}^-$, it has been argued that 
couplings derived from density functional theory (DFT) are too weak \cite{PRL1995.Gunnarsson}.
Contrary, DFT couplings were later shown to accurately account for photoemission 
data on C$_{60}^+$ \cite{PRL2003.Manini}. This apparent controversy calls for
complementary studies of the e-ph interaction that put DFT-derived couplings to test.

  \begin{figure}[t]
    \includegraphics[width=.88\columnwidth]{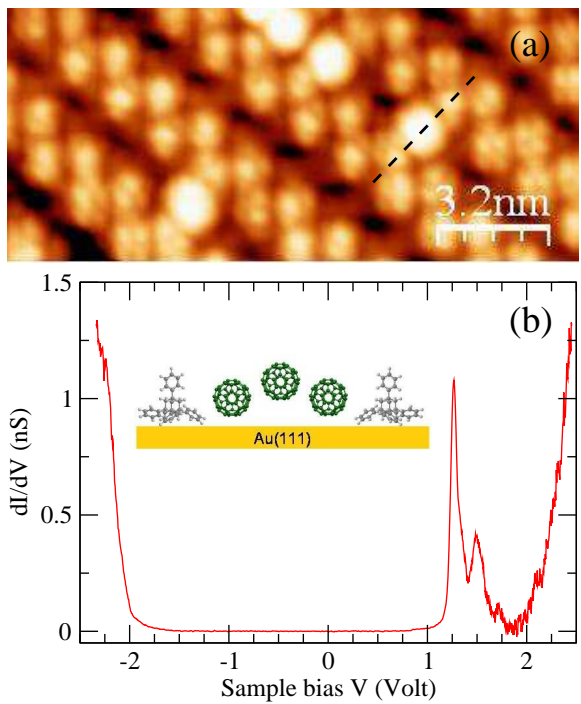}
    \caption{(color online) 
      (a) STM image of the double rows of alternating TPA/C$_{60}$ on Au(111) ($I$ = 16 pA, $V$ = 580 mV, $T$ = 5 K). 
      Some isolated C$_{60}$ molecules are adsorbed in the troughs between the rows. 
      (b) Side view of the structural model along the dashed line in (a) and 
      dI/dV spectra recorded on top of such a fullerene 
      ($I_\mathrm{set}$ =  0.46 nA,  $V_\mathrm{set}$ = 2.5 V, $V_\mathrm{rms}$=14 mV).
   } 
    \label{fig:1}
  \end{figure}

In this Letter, we report on the JT fingerprint in the conductance spectra of single C$_{60}$ molecules
in a tunneling junction. Fullerenes were decoupled from a Au(111) surface by a template of organic molecules. 
Scanning tunneling spectroscopy reveals a broad sideband above a sharp LUMO-derived resonance. To understand this we calculate the 
tunneling spectrum by solving the JT problem from first principles. The vibrational modes, frequencies, and e-ph couplings 
are derived from DFT, and the excitation spectrum of the \emph{vibronic} ground state is computed numerically by 
nonequilibrium Green's function (NEGF) techniques as well as by exact diagonalization. 
The theoretical results explain the origin of the experimentally observed fine structure as due to the dynamic JT effect.

The electronic decoupling of the molecule from its metallic substrate is crucial for studying any JT-induced
fine structure of the molecular resonances. To reduce this interaction experimentally we use a novel strategy based on 
codeposition of C$_{60}$ and 1,3,5,7-tetraphenyladamantane (TPA) on Au(111), which results in spontaneous formation of 
nanostructures where the C$_{60}$ cages are lifted away from the surface by support from TPA molecules \cite{PRL2008.Franke,
JPCondMat2008.Torrente}. One of the observed structural motifs are double rows of alternating TPA and C$_{60}$, cf.~Fig.~\ref{fig:1}(a).
This dielectric template yields troughs for single C$_{60}$ molecules to be trapped between neighboring 
double rows as sketched in Fig.~\ref{fig:1}(b). 
Differential conductance ($dI/dV-V$) spectra taken over the center of these molecules 
reveal strong non-linearities around $-2.2$ V, 1.3 V and 2.5 V, associated with the HOMO, LUMO and LUMO+1 
derived resonances, respectively (Fig.~\ref{fig:1}(b)). The corresponding large gap of $\sim$ 3.5 eV is a clear sign 
of a weakly interacting molecule with its surrounding \cite{JPCondMat2008.Torrente}.
Even more striking is the observation of a very sharp LUMO-derived resonance (FWHM $\sim$ 60 meV),
reflecting a long lifetime in the order of 10 fs of the tunneling electron. 
Hence, this system is ideal to study e-ph couplings. However, instead of single vibronic peaks \cite{JPCB2005.Pradhan}, 
we observe a broad sideband at $\sim$ 230 meV above the LUMO energy, which lies well outside of the 200-meV-wide vibrational
spectrum of C$_{60}$.

Tunneling electrons through an isolated C$_{60}$ molecule will probe the excitation spectrum 
as resulting from a dynamic JT interaction with the vibrations. Near the LUMO-derived resonances
in focus here, we exclude the HOMO and LUMO+1 states from our treatment since these 
are well separated in energy. Further, group theory gives that only the $A_g$ and $H_g$ 
intramolecular phonons couple to the $t_{1u}$ LUMO states \cite{Manini-Dunn}.
Thus, the JT problem for C$_{60}^-$ is described by the following Hamiltonian
\begin{eqnarray}
H &=& \varepsilon_0\sum_{i=1}^3 c_i^\dagger c_i + \sum_{\nu=1}^{42} \hbar\omega_\nu b_\nu^\dagger b_\nu \nonumber\\
 &&+\sum_{\nu=1}^{42}\sum_{i=1}^3\sum_{j=1}^3 M_{i,j}^\nu c_i^\dagger c_j(b_\nu^\dagger+b_\nu),
\end{eqnarray}
where $c_i^\dagger$ ($c_i$) is the one-electron creation (annihilation) operator corresponding to one of the 
three degenerate single-particle LUMO states $|i\rangle$ and $b_\nu^\dagger$ ($b_\nu$) is the bosonic creation (annihilation) operator 
of each of the 42 (8 fivefold degenerate $H_g$ and 2 non-degenerate $A_g$) vibrational modes $\nu$. 
The parameters $\varepsilon_0$, $\omega_\nu$, and $M_{i,j}^\nu$ represent the bare LUMO energy, 
the vibrational frequency of mode $\nu$, and the coupling constant for
the scattering of an electron from state $|i\rangle$ to state $|j\rangle$
under creation or annihilation of a phonon in mode $\nu$, respectively. These
parameters (see Tab.~\ref{tab:1}) are obtained from DFT calculations \cite{CalcDetail,soler.02.siesta,noteB} 
applying the scheme described in Ref.~\cite{PRB2007.Frederiksen} to the neutral \cite{PRB2002.Saito}, isolated C$_{60}$ molecule.

  \begin{table}[t]
    \begin{tabular}{ c c c p{0mm}| c c c}
      \hline \hline
      Mode     &  $\hbar\omega_\nu$ & $\lambda_\nu/N(0)$ && Mode     &  $\hbar\omega_\nu$ & $\lambda_\nu/N(0)$\\
      $\nu$  & meV (cm$^{-1}$) & meV                && $\nu$    &  meV (cm$^{-1}$)   & meV\\
      \hline
      $H_g$(8) & 195 (1572) & 14.6                  && $H_g$(4) &  95 (766) &  4.0\\
      $A_g$(2) & 185 (1491) &  7.3                  && $H_g$(3) &  86 (693) & 10.2\\
      $H_g$(7) & 178 (1439) & 15.0                  && $A_g$(1) &  60 (484) &  1.0\\
      $H_g$(6) & 155 (1251) &  3.2                  && $H_g$(2) &  52 (419) & 11.6\\
      $H_g$(5) & 136 (1094) &  4.3                  && $H_g$(1) &  32 (256) &  4.7\\
      \hline \hline
    \end{tabular}
    \caption{Partial electron-phonon coupling constants $\lambda_\nu/N(0) = 2/9\sum_{i,j}|M_{i,j}^\nu|^2/\hbar\omega_\nu$ of 
      the $A_g$ and $H_g$ intramolecular modes for C$_{60}$, c.f.~Eq.~(10) of Ref.~\cite{PRB1993.Antropov}, with $N(0)$ being the
      density of states. The sum of all modes gives a coupling strength of $\lambda^\mathrm{GGA}/N(0)=76.1$ meV \cite{noteB}
      in good agreement with previous calculations \cite{PRB1993.Antropov,RMP1997.Gunnarsson,PMB2001.Manini,PRB2002.Saito}.}
    \label{tab:1}
  \end{table}

In our STM configuration the molecule under investigation is much less
coupled to the tip than to the substrate, i.e., $\Gamma_\mathrm{t}\ll\Gamma_\mathrm{s}$ in terms of the tip $\Gamma_\mathrm{t}$
and substrate $\Gamma_\mathrm{s}$ tunneling rates. Under this condition the electron occupancy of 
the molecular states is effectively in equilibrium with the substrate and the differential conductance at a bias voltage 
$V_\mathrm{s}$ is to a first approximation proportional to $\rho(\mu_\mathrm{s}+eV_\mathrm{s})\Gamma_\mathrm{t}$,
where $\rho$ is the local density of states (LDOS). This takes into account both couplings to the leads as well as 
to the vibrations. We have considered two 
different methods for calculating $\rho$, namely exact diagonalization with the Lanczos scheme (LS) as well as NEGF within
the self-consistent Born approximation (SCBA) \cite{noteC}.

  \begin{figure}[t]
    \includegraphics[width=.95\columnwidth]{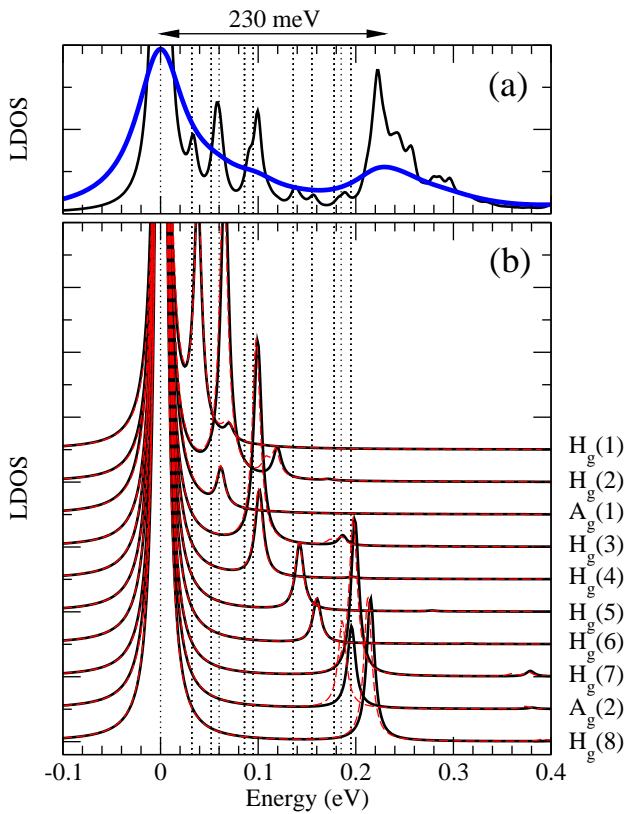} 
    \caption{(color online). (a) Theoretically computed JT-induced fine structure in the LDOS
      of the LUMO via the coupling to $A_g$ and $H_g$ modes, shown for two different intrinsic level broadenings: 
      $\Gamma =$ 10 meV (black line) and $\Gamma =$ 60 meV (thick blue line).
      (b) Analysis of the position of the phonon sidebands including only one mode at the same time, calculated with both
      the SCBA (black lines) and LS (dashed red lines) methods ($\Gamma =$ 10 meV). For clarity the spectra are off-set and
      shifted to place the main peak at zero energy. 
      The vertical dotted lines indicate the fundamental frequencies of the $A_g$ and $H_g$ modes. }
    \label{fig:2}
  \end{figure}

Figure~\ref{fig:2}(a) shows the computed JT spectrum from SCBA for two different values of the total
electronic broadening $\Gamma=\Gamma_\mathrm{t}+\Gamma_\mathrm{s}\approx\Gamma_\mathrm{s}$ (FWHM). 
The spectrum is composed of a main peak followed by a series of sidebands at higher excitation energies. Depending on the
broadening, the LUMO spectrum might appear as one main peak with a single sideband separated by 230 meV. A similar structure
has also been reported in photoemission \cite{PRL1995.Gunnarsson,CPL1997.Bruhwiler,PRL2002.Canton,PRL2003.Manini}. 
It is important to realize that the fine structure in the LDOS does
not relate in a simple way to the fundamental frequencies of the $A_g$ and $H_g$ modes (in the range 32-195 meV as indicated
with vertical dotted lines in Fig.~\ref{fig:2}). Hence, for JT systems it is not possible to directly relate
vibronic structure in spectroscopy with molecular frequencies. Another important fact is that the
vibronic spectrum is threefold degenerate, i.e., the parent symmetry of the LUMO is restored by the $A_g$ and $H_g$ 
modes as expected from the \emph{dynamic} JT effect \cite{noteD}. Hence, the molecule is not statically distorted.

  \begin{figure}[t!]
    \includegraphics[width=.85\columnwidth]{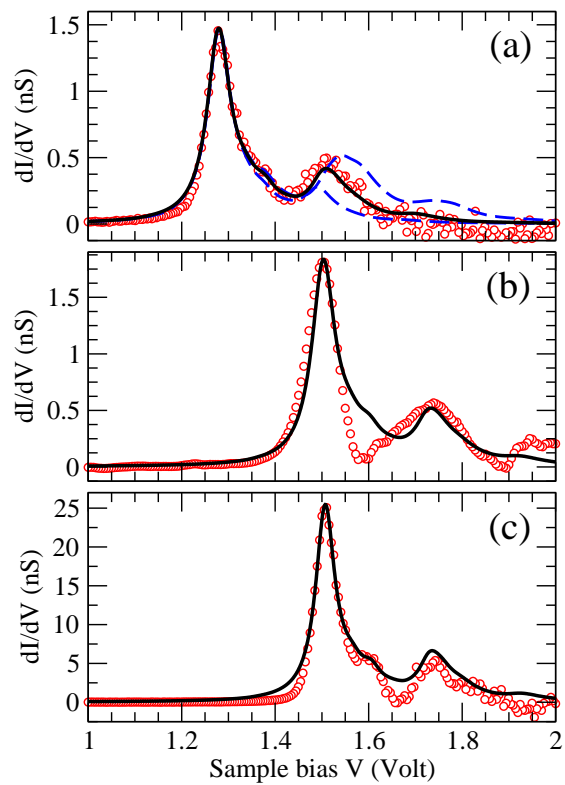} 
    \caption{(color online). Direct comparison between experiment (red circles) and theory (black line) for three different molecules:
      (a) $I_\mathrm{set}$ =  0.3 nA,  $V_\mathrm{set}$ = 2.5 V, $V_\mathrm{rms}$ = 7 mV, $\Gamma_\mathrm{s}$ = 60 meV.
      (b) $I_\mathrm{set}$ =  0.52 nA,  $V_\mathrm{set}$ = 2.5 V, $V_\mathrm{rms}$ = 3.5 mV, $\Gamma_\mathrm{s}$ = 60 meV.
      (c) $I_\mathrm{set}$ =  1.2 nA,  $V_\mathrm{set}$ = 2.5 V, $V_\mathrm{rms}$ = 7 mV, $\Gamma_\mathrm{s}$ = 50 meV.
      Panel (a) includes calculations with scaled e-ph coupling constants (dashed blue lines) corresponding to 
      $\lambda_\nu\rightarrow 1/{2}\,\lambda_\nu^\mathrm{GGA}$ and $\lambda_\nu\rightarrow {2}\,\lambda_\nu^\mathrm{GGA}$.
    }
    \label{fig:3}
  \end{figure}

To gain an understanding of the contributions from the different vibrational modes to the total spectrum, 
we show in Fig.~\ref{fig:2}(b) the LDOS as resulting from a calculation with the coupling to only one type of mode at 
a time. Although the sideband structure is much simpler, one observes that the peak separation for all $H_g$ modes
is larger than the phonon energy. This nontrivial behavior is generic to the $T_{1u}\otimes h_g$ JT problem
\cite{Manini-Dunn}. Figure \ref{fig:2}(b) also shows that, except for the more strongly coupled $A_g(2)$ mode,
SCBA and LS yield essentially the same results for the LDOS due to the weak e-ph couplings.

We next turn to a comparison with the experiment. The measured tunneling spectra on different molecules
exhibit certain variations in peak position and structure around the LUMO resonance. This variation is likely
due to slightly different environments, and possibly also molecular orientation, of the selected 
C$_{60}$ species. In Fig.~\ref{fig:3} we compare the $dI/dV$-spectra of three different molecules
with theoretical spectra by applying an appropriate substrate coupling $\Gamma$.
The relative peak height as well as the position of the sideband with respect 
to the main peak are in very good agreement. This fact supports the interpretation that 
these spectra indeed exhibit the free-molecule JT effect.

Alternative mechanisms, such as the weak spin-orbit coupling in carbon \cite{dresselhaus.95} or the Stark 
effect from the applied voltage, cannot account for the 230 meV splitting seen in Fig.~\ref{fig:3}.
Although details in the weak electronic coupling of the C$_{60}$ molecule to 
its environment may have an influence in the experimentally recorded spectra, we believe
that this effect solely cannot produce such an agreement as the JT theory presented above for a decoupled
molecule. Furthermore, from the different timescales of electron tunneling events and residence times in 
the molecule, we can also exclude heating effects. Based on the NEGF-SCBA calculations we find that 
heating would be important if the current through the LUMO states is raised by two orders of magnitude.

In Fig.~\ref{fig:3}(a) we also show how the calculated spectrum changes by a scaling of the DFT-derived e-ph couplings
corresponding to $\lambda_\nu\rightarrow 1/{2}\,\lambda_\nu^\mathrm{GGA}$ and $\lambda_\nu\rightarrow {2}\,\lambda_\nu^\mathrm{GGA}$.
As seen, these scalings decimate the agreement with experiment in terms of peak ratio and separation, thus pointing
towards reasonable values of the e-ph coupling strength within DFT.

In summary, we have explained the dynamic JT effect as it is revealed in electron tunneling spectroscopy through the 
C$_{60}^-$ anionic state of sufficiently isolated molecules. The experimental realization of this situation was based on
a novel experimental preparation procedure of codeposition of C$_{60}$ and TPA molecules on Au(111), where C$_{60}$ molecules 
on top of the double row template are exceptionally decoupled, geometrically and electronically, from the metallic substrate.
Low-temperature STM $dI/dV$ spectra recorded under such conditions display a delicate fine structure around the LUMO-derived
resonance, that closely resembles the theoretically computed JT spectrum. This quantitative agreement further supports 
that the calculated e-ph couplings from DFT provide an accurate description for the vibrational interactions 
in the molecule.

TF acknowledges support from the Danish FNU (grant 272-07-0114).
AA and TF thank partial financial support from UPV/EHU (grant IT-366-07) and MEC (grant FIS2007-66711-C02-01).
NL acknowledges support from MEC (grant FIS2006-12117-C04-01).
Financial support by the DFG through SPP 1243 and Sfb 658 is gratefully acknowledged.
We thank S.~Zarwell and K.~R\"uck-Braun for the synthesis of the TPA molecules.
JIP and KJF acknowledge fruitful discussions with F. von Oppen.



\begin{thebibliography}{nnnyys}
\bibitem{noteA} The only exception being ``accidental'' degeneracy, i.e., when the electronic
         degeneracy not caused by symmetry.
\bibitem{CalcDetail} We use the SIESTA implementation of DFT \cite{soler.02.siesta} with an optimized 
         double-$\zeta$ plus polarization basis set for the valence electrons, 
	 the generalized gradient approximation (GGA-PBE) for exchange-correlation, 
	 a relativistic Troullier-Martins pseudopotential for the core electrons, 
	 and a 200 Ry cutoff for real-space grid integrations.
\bibitem{noteB} The local density approximation (LDA) leads to a similar
         coupling strength $\lambda^\mathrm{LDA}/N(0)=74.2$ meV, but vibrational frequencies are generally 
	 overestimated.
\bibitem{noteC} The polaronic Hartree diagram is neglected.
\bibitem{noteD} As expected from the JT theorem, the electronic 
         degeneracy is broken if the symmetry of the e-ph couplings are disturbed, 
	 e.g., by coupling only to one of the five $H_g(8)$ modes.
\end{thebibliography}
\end{document}